# Zigzag Filamentary Theory of Broken Symmetry of Neutron and Infrared Vibronic Spectra of YBa$_2$Cu$_3$O$_{6+x}$

J. C. Phillips[1] and J. Jung[2]

1. Bell Labs., Lucent Tech. (Retired),  Murray Hill, N. J. 07974-0636
2. Dept. Physics, Univ. of Alberta, Edmonton, AB  T6G 2J1, Canada

Filamentary high-temperature superconductor (HTSC) theory differs fundamentally from continuous HTSC theories because it emphasizes self-organized, discrete dopant *networks* and does not make the effective medium approximation (EMA).  Analysis of neutron and infrared (especially with c-axis polarization) vibrational spectra, primarily for YBa$_2$Cu$_3$O$_{6+x}$, within the filamentary framework, shows that the observed vibronic anomalies near 400 cm$^{-1}$ (50 meV) are associated with curvilinear filamentary paths. These paths pass through the cuprate chains and planes, as well as resonant tunneling centers in the BaO layers.  The analysis and the data confirm earlier filamentary structural models containing ferroelastic domains of diameter 3-4 nm in the CuO$_2$ planes; it is these *nanodomains* that are responsible for the discrete glassy nature of both electronic and vibrational properties.  Chemical trends in vibronic energies and oscillator strengths, both for neutron and photon scattering, that were anomalous in continuum models, are readily explained by the filamentary model.

## I.    INTRODUCTION

Apart from their high superconductive transition temperatures, the layered pseudoperovskite cuprates exhibit many other anomalous properties.  The most important may be the many anomalies in the temperature and frequency dependencies of their normal-state transport properties.  Often linear temperature dependencies are observed that are quite different from the quadratic variation expected in normal metals described by Fermi liquid theory;  the composition dependencies of these linearities correlate well with that of T$_c$.  This has led to the idea that the superconductive intermediate phase should be described as a "non-Fermi liquid", a popular, but awkward, fuzzy and uninformative, phrase that recognizes the existence of the anomalies without explaining their origin.

A major question for any microscopic theory of high temperature superconductivity (HTSC) is whether electron-phonon (e-p) interactions are responsible for HTSC, just as has been shown[1] for metallic superconductors. This question has proved to be extremely difficult to answer, because of the complexity of the internal structure of HTSC. For example, the proof for the old metallic superconductors rested largely on observation of an isotope effect. A sizable isotope effect, comparable in magnitude to that predicted by the Bardeen-Frohlich interaction, is indeed observed[2] for $YBa_2Cu_3O_{6+x}$ near the metal-insulator transition at $x = 0.4$, but the Cu and O shifts have opposite signs. Moreover, as x increases to its optimal value 0.9 where $T_c$ is maximized, both isotope shifts become small.

It is possible that phonons, especially longitudinal optic (LO) phonons, that also exhibit many anomalies, can be used as excellent microscopic probes of the internal structures and mechanisms of HTSC. If strong electron-phonon interactions are present, the current-carrying wave packets may be best described as polarons or vibrons, that is, wave-packet products of electronic excitations and phonons, especially LO phonons, that interact most strongly with electronic excitations because longitudinal interactions are largely unscreened.

It has been argued[3] that these anomalies, as well as many others, arise from the breakdown of the effective medium approximation (EMA). (Fermi liquid theory is the simplest metallic realization of the EMA.) There is a growing awareness among experimenters that *spatial inhomogeneities* play an important part in the anomalous properties of HTSC, as stressed very early by Mueller et al[4]. However, this phrase, while correct, also does not constitute a satisfactory microscopic theory. Moreover, analogies with spin density waves (SDW), charge density waves (CDW), stripes, etc., are not particularly helpful, because such excitations, involving long-range order, correspond to alternative pair scattering channels, and thus have long been known to be *complementary*, or destructive, to Cooper pairing and superconductivity[5]. Of course, one could begin with the EMA and attempt to refine it by adding local field corrections, in the spirit of Lorenz, but superconductive phase coherence is an extensive (global) property, and this approach, based on entirely incoherent local refinements, is probably inadequate.

The correct approach, in the authors' view, is to develop a global topological model that includes the local topologies that are expected in perovskite and pseudo-perovskite materials characterized by *strong ferroelastic forces*. The filamentary model, first discussed[6] more than 10 years ago, is such a global topological model. It enables us to study the effects of intralayer ferroelastic planar partitioning into nanodomains, and interlayer coupling through dopant resonant tunneling centers, on the phase diagram[7], for example. The model explains why the first transition from the underdoped phase to the intermediate optimally doped, or filamentary, phase is *continuous*. It also explains why, in very carefully prepared samples, the second transition from the filamentary phase to the overdoped, or Fermi liquid, phase is *first order*. It shows that these strongly asymmetric transitions in the phase diagram are essentially topological, and although strongly affected by many-electron Coulomb interactions, are not essentially dependent on them. Such asymmetric, self-organized dopant intermediate phases have never been observed in connection with long-range order due to SDW, CDW, or stripes, and there is no apparent reason why they should have been observed.

Similar topological effects have, however, been observed in intermediate, self-organized phases of strongly disordered network glass alloys that can be described by classical short-range force (spring) models[8]. They appear to be characteristic of *connectivity transitions* (molecular or atomic) in flexible, self-organized disordered materials. Because the full characterization of filamentary effects in disordered *quantum* systems may be impossible (it appears to be equivalent to constructing a very large quantum computer[9]), these analogies to a topologically isomorphic *classical* system are very important. In particular, the effects of self-organization in network glasses have been successfully simulated numerically[10]. The experimental effects described above (the EMA single continuous transition splits into two transitions; with increasing doping or connectivity, the first transition is continuous and the second is first order) are actually obtained in these simulations, which are illustrated for the reader's convenience in Figs. 1 and 2. The correspondence principle suggests that there is every reason to believe that the same effects would happen in simulations of glassy network quantum systems, if we could only simulate phase coherence with a very large quantum computer programmed to include the effects of many-body (electron and phonon) self-organization.

It is important to emphasize here that in the absence of discrete dopant self-organization, there would be no intermediate phase. The transition from the insulator to the Fermi liquid would be continuous, and there would be no HTSC. Although filamentary self-organization is not associated with long-range order, and so is not observable by diffraction, it creates characteristic features in the neutron and infrared spectra that cannot be explained by continuum models. This paper examines some of these features, and it shows that they behave qualitatively (and in some cases semi-quantitatively) as one expects from a filamentary model. By and large, continuum models are not only unable to account for the main features of the phase diagram, but they also cannot explain the anomalies in the neutron and infrared spectra discussed here.

## II.     GENERAL FEATURES OF NEUTRON AND INFRARED DATA

The first indications that conventional models based on the EMA are structurally inadequate to describe HTSC arose in an unexpected way. If we look at the neutron data[11-13], the Raman and infrared data[14], and model force-constant fits[11,14] to these data on $YBa_2Cu_3O_{6+x}$, we can see how this happened. The neutron data show pronounced differences between the $Y-O_6$ and $Y-O_7$ data (Fig. 3(a)), and these differences cannot be accounted for with simple spring-constant models by the loss of only one O from the chains[11]. The situation is even worse for the $Y-(Cu_{0.9}Zn_{0.1})_3-O_7$ data (very similar to $Y-O_6$) compared to the $Y-Cu_3-O_7$ data (Fig. 3(b)), where the number of atoms is the same, and the spring changes should be much, much smaller. The most important difference is the enhanced "effective" density of states $G(\omega)$ in the region 40 meV $\leq \omega \leq$ 60 meV, with the growth of a TO peak near 40 meV, labeled A, and the appearance of an entirely new peak B near 50 meV. Note also that for Y-O$_7$ there is a similar enhancement (about 1/3 as large) on going from T = 92K to T = 11K. These differences will be explained here as the result of enhanced filamentary coherence in the superconductive state. In other words, a selected subset of Y-O$_7$ phonons in the region 40 meV $\leq \omega \leq$ 60 meV scatters neutrons much more coherently than the Y-O$_6$ phonons. Such selectively coherent scattering of neutrons by phonons is novel, but it is necessary to explain the data shown in Figs. 3(a,b).

One might think that infrared studies of optical phonons would be of little value in constructing filamentary models of HTSC, because such optical data could always be interpreted within the EMA framework with a sufficiently large number of parameterized Lorentzian oscillators. However, here again anomalies near 50 meV have been found in YBa$_2$Cu$_3$O$_{6+x}$, especially for **E**∥c, that have puzzled experimenters for more than five years[15,16]. The main anomaly is the development of a very broad electronic (vibronic) band $M_1^c(x)$ in $\sigma^c$ between 50 and 70 meV (400 cm$^{-1}$ and 550 cm$^{-1}$) (labelled "?") that becomes more intense at low doping and low temperatures, as shown in Fig. 4. This band disappears with Zn doping as superconductivity itself does[17], just as in the neutron data, and it borrows oscillator strength not only from transverse optical (TO) apical- and planar oxygen *ab* modes, but also from LO chain modes and the non-Drude *ab* background conductivity. The temperature-dependent growth of oscillator strength of this c-axis polaronic band is a larger effect than the change in "effective" neutron density of states G($\omega$), but its spectral dependence is similar. The larger effect for photons is expected, and can be attributed to longer filamentary coherence length for the electronic part of the vibronic wave function.

It should be stressed that the x-dependence of the $M_1^c$ band is somewhat different from that of the extra 50 meV peak in G($\omega$); this will be explained also in terms of differences in coherence effects for neutrons and photons. Because it involves defects such as O vacancies, it is also quite sensitive to the details of the procedure used to grow the single crystals, and thus some of our discussion focuses on the data recently obtained by reflection from shiny as-grown surfaces of crystals grown in special crucibles[18].

Tunneling centers in the infrared absorptive surface layers of these crystals are probably more completely self-organized than in the surfaces of samples which have been damaged by mechanical polishing. In general, in the EMA such sensitivity to surface preparation is unlikely to be characteristic of the bulk crystal. Thus the observed large differences in frequency and temperature dependence of the $M_1^c$ band already provide substantial support for a microscopic model in which defects, or tunneling centers, play a critical role.

For a long time the origin of the $M_1^c$ band was frankly described[15] as a mystery related to some kind of broken symmetry, but recently it has become fashionable to ascribe it to an "infrared *c*-axis Josephson" effect[19,20]. In practice this means that in a one-dimensional model one adds two featureless, broad-band absorptive background terms, of unknown structural origin, to the phonon dielectric function[20]. The parameters of the first term are adjusted to describe the pseudogap, while the parameters of the second term are adjusted to fit the $M_1^c$ band. No microscopic justification is offered for either term, and neither is able to explain observed chemical trends without suitable further parameter adjustment.

From his experience with the microscopic theory of superconductive tunneling[21] that was used by Josephson to derive his dc effect[22], one of the authors finds little physics in the phrase "Josephson plasmon" beyond the kind of slack metaphor or fuzzy analogy that was used in the context of the so-called[23] RVB theory[24] of HTSC. A plasmon is a charge *density* fluctuation that does not involve the pair amplitude Josephson *phase*. The combination of these two mutually inconsistent (oxymoronic) terms is an extreme example of metaphorical physics, so extreme as to be suggestive of metaphysics. Indeed, most recently there has been some retrenchment: it has been remarked[25] that this "Josephson" analogy has been used mainly to describe local-field effects, and that a more appropriate phrase would involve interlayer "plasmons" [of unknown origin] only.

Another grave weakness of the "Josephson plasmon" model of the infrared anomalies is that it does not explain how very similar anomalies can be produced by such plasmons in the effective vibrational density of states $G(\omega)$ measured by neutrons, as no mechanism for coupling neutrons to plasmons is known. It would seem that this rather basic observation eliminates interlayer Josephson plasmons[18-20] as a possible mechanism[24] for HTSC.

Here it will be shown that, in spite of its limitations as only a global topological model, filamentary theory, which has already explained microscopically the anomalous linear temperature dependencies of conductivities, and the origin of the pseudogap, provides a much more quantitative description of the origin of the $M_1^c$ band. The band arises in part because of *broken symmetry*, as proposed in the presentation of the original data[26], and involves absorption by phonons that would be described as TO-LO hybrids in the EMA. This interpretation appears to have been excluded recently, because LO phonons have zero oscillator strength in the EMA. Thus the long-delayed *confirmation* of the broken symmetry of this band may help to dramatize the fundamentally inadequate nature[3] of the EMA (and, more generally, metaphorical physics, or metaphysics) as a platform for microscopic theories of HTSC.

### III. BROKEN SYMMETRY: INFRARED LO PHONON SELECTION RULE

It would seem obvious that LO phonons cannot couple to transverse electromagnetic radiation, and this selection rule has been proved many times in cubic III-V and II-VI semiconductive materials for which the EMA is valid. In those materials both the electronic and vibrational states have the modulated plane-wave character of Bloch states. The modulation functions do not alter the symmetry selection rules. In particular, while the LO phonons at $\mathbf{q} = 0$ identified by neutron scattering have zero infrared oscillator strength, those frequencies are associated with zeroes in $\varepsilon_1(\omega)$ that give rise to maxima in the dielectric loss function $\text{Im}(-1/(\varepsilon_1 + i\varepsilon_2))$.

That the symmetry of the EMA is strongly broken in HTSC is shown by excellent heuristic correlations[26] between minima in planar $\varepsilon_2^{ab}$ (or $\sigma^{ab}$) and maxima of axial $\text{Im}(-1/\varepsilon^c)$ for many cuprates (YBCO, $Pb_2Sr_2DyCu_3O_8$, $Bi_2Sr_2CaCu_2O_8$, and $Tl_2Ba_2CaCu_2O_8$). An example[26] of this correlation, based entirely on measured dielectric functions, is shown, for the reader's convenience, for YBCO$_7$ in Fig. 5; note the minimum in $\sigma^{ab}$ and maximum of Im($-1/\varepsilon^c$) near $\omega$ = 430 cm$^{-1}$. [This is much lower than the frequency 515 cm$^{-1}$ of the zone center planar LO phonon predicted by a shell model[14]. This disagreement, which is very large (50% in $\omega^2$) for an optic mode frequency, can be explained as a result of filamentary coherence effects omitted in the shell model, which assumes an ideally ordered lattice structure. However, it does agree with the measured frequency of the *ab* zone-boundary LO phonon.]

One could argue that this correlation is accidental: by doing so, we will see why it is not. The maximum of Im($-1/\varepsilon^c$) occurs when $\varepsilon^c(\omega) = 0$. Such a node in the *c*-axis dielectric function is produced after a strong absorption band associated with a *c* axis TO phonon. The frequency shift $\omega^2_{LO} - \omega^2_{TO}$ depends on the ionic polarizability. The latter should be only slightly smaller along the *c* axis than in the *ab* plane. Thus one could argue that the coincidence is expected.

The weakness of this dismissive correlation with the minima in $\sigma^{ab}$ is that it does not explain the origin of the $M_1^c$ peak. This peak was, however, discovered by following the implications of the original heuristic correlations, just as HTSC was discovered by looking for materials with ferroelastic (very strong electron-phonon) interactions. The $M_1^c$ peak appears most strongly at low T centered near 380 – 430 cm$^{-1}$ for x = 0.5-0.6 in early neutron data[11,12], and less strongly near 410 cm (x = 0.5) and 480 cm (x = 0.75) in more recent c-axis infrared data[18] on better samples with enhanced peak intensities. This broad vibronic peak is much more sensitive to sample and surface quality than the narrow phonon peaks. Its frequency can also be compared to that of chain LO phonons observed by neutron scattering[27]. These frequencies range from ~ 580 cm$^{-1}$ near the planar zone center, **q** = 0, to ~ 450 cm$^{-1}$ near the planar zone edge. In the context of polarons localized near a resonant tunneling center, the zone edge states may form more localized wave packets, so that the latter value is the one that could be compared to the frequency of the $M_1^c$ peak. The broad non-Drude background absorption at higher energies[25] is then associated with excitation of a screening cloud of polarons localized in nanodomains.

At this point one of the advantages of the filamentary model, compared to the EMA "Josephson plasmon" model, becomes quite obvious. Instead of having to postulate the existence of two absorption bands of essentially unknown origin, we can see that all we need to be able to do is to understand the origin of the electron-phonon coherence and why the symmetry of the LO phonons has been broken[26]. Such coherence and broken symmetry occur naturally in the 1990 filamentary model[6], and are indeed one of its characteristic features, so that no new assumptions or metaphorical constructs (internally consistent or not) are necessary; in other words, we immediately have very substantial economy of means.

Oscillator strength transfer between the $M_1^c$ peak and TO phonons is closely associated with superconductive order and is thus an effect that depends on phase coherence. The temperature dependence of the oscillator strength[18] of the $M_1^c$ peak is compared to the oscillator strength *lost* by the *c* oxygen bond-bending TO mode at 320 cm$^{-1}$ in Fig. 6, reproduced here for the reader's convenience. The overall strength of the $M_1^c$ peak is about two-three times larger than the oscillator strength *lost* by the oxygen bond-bending mode, but the temperature dependencies are quite similar. What this means is that the strength of the $M_1^c$ vibronic peak can be borrowed from four sources: the *c* planar and apical oxygen bond-bending TO vibrational modes, from the LO chain vibrational mode, and from the *ab* non-Drude background electronic conductivity, as seen in Fig. 5. It seems obvious that the broken symmetry associated with this vibronic transfer of oscillator strength cannot be explained by any Fermi liquid electronic model based on plane-wave basis states.

The zigzag filamentary model shows geometrically how this extensive sharing of oscillator strength happens, as an unavoidable result of the three-dimensional "maze" structure of the insulating barriers formed by planar nanodomains and interplanar resonant tunneling centers. It also explains the origin of the non-Drude background electronic conductivity, and shows why four different vibrtional and electronic oscillator strengths combine to form the vibronic $M_1^c$ band. Finally, the vibronic coherence that has already been discussed[28] in the context of LO phonon dispersion normal to the chains and intensity anomalies in neutron scattering data accounts for the rapid growth of the integrated intensity of the $M_1^c$ peak below $T_c$.

## IV. ZIGZAG FILAMENTARY MODEL: BROKEN SYMMETRY AND SHARP DEFLECTIONS

In Fig. 7(a) the zigzag filamentary model is illustrated schematically for YBCO in a cross-section normal to the *a* axis. Only three layers are shown schematically: I, a "metallic" $CuO_2$ layer, a semiconductive BaO layer, and II, a *b*-axis $CuO_{1-x}$ chain segment. The key structural elements omitted in EMA models are the *semiconductive* nanodomain walls in the "metallic" $CuO_2$ layers I, the vacancies in the $CuO_{1-x}$ chain segments, and the dopants in the semiconductive layers. These $CuO_2$ layer domain walls are often, but not necessarily always, antiferromagnetic, and they form the pattern of an *irregular* (glassy) metal/semiconductor chessboard. The forces that drive the formation of the glassy $CuO_2$ layer nanodomain structure are not the antiferrromagnetic ones found in *both* oxides and halides, but are instead the ferroelastic ones which are *characteristically large* in all perovskites and related oxides (two-fold coordinated anions), but not in halides (one-fold coordinated anions). Unlike antiferromagnetic interactions, which become weak above the Neel temperature, these much more *perovskite-specific* ferroelastic oxide interactions are very strong and persist up to the melting point, and are thus able to explain the normal state transport anomalies[7], which persist up to the annealing temperature ~ 700K. It was the existence of such strong lattice interactions in perovskite-like oxides that motivated the discovery of HTSC[29].

In the zigzag percolative model the dopants, which are located in the semiconductive layers, are assumed, after annealing, to develop short-range order relative to the nanodomains in the $CuO_2$ layers by forming *percolative metallic filaments* that are optimized to maximize the conductivity. Optimization by selective dopant ordering occurs when the sample conductivity is maximized at high temperatures, as the formation of these filaments reduces the free energy by improving the screening of internal ionic electric fields. The dopants act as resonant tunneling centers that strongly couple the metallic $CuO_2$ nanodomains to the $CuO_{1-x}$ chain segments to form the zigzag filaments. The nanodomain configuration is irregular because of grown-in stresses and composition fluctuations, so the optimized dopant configurations are also irregular and have an intrinsically glassy nature.

In the zigzag percolative model the filamentary paths are composed of structural units based on three path segments: I, metallic $CuO_2$ nanodomains, II, $CuO_{1-x}$ chain segments, and III, interplanar resonant tunneling centers (RTC's). These alternate doubly as I, III, II, III, I, III, etc. One can now ask, which of these three structural elements is most easily identified as a chemical trend or difference term in the phonon spectra, measured either by infrared or by neutrons?

The metallic $CuO_2$ nanodomain contribution I may be identified as the much-discussed softening of the metallic $CuO_2$ TO mode with decreased oxygen doping of the chains, see Fig. 5 of ref. 15. The interplanar resonant tunneling center (RTC) contribution will be difficult to identify, because vibrons at RTC's are so localized that they will include an admixture of many phonon modes. The $CuO_{1-x}$ chain segments III should be identified most easily, because they are essential in forming bridges that enable the polarons to bypass $CuO_2$ nanodomain walls. Moreover, current flowing along $CuO_{1-x}$ chain segments III can entrain strongly with LO phonons localized on III, which increases the scattering strength for one-phonon LO III processes.

Softening effects for $CuO_2$ TO modes have been extensively discussed in the diffraction literature[30], but the evidence for strong LO III phonons seems to have been largely ignored. This evidence is already strong in the neutron and Raman data[11-14], as discussed in Sec. II and Fig. 4. Further insight is gained from the neutron spectra[31] of Y-$O_{6.5}$ and $La_{2-x}Sr_xCuO_4$ for x = 0 and 0.15. These show a close similarity near 400 cm$^{-1}$ between Y-$O_{6.5}$ and Y-$O_6$. Thus the new peak and enhanced structure grows most rapidly in the neutron data for Y-O$_x$ probably near x = 6.75, the center of the ortho I-ortho2 transition where T$_c$ varies most rapidly. In LSCO the LO peak is at 85 meV, and the new peak appears near 70 meV, which is much closer to the LO peak than in YBCO. It follows that the new peak at 50 meV in YBCO cannot be associated only with the CuO$_2$ planes that are already present in LSCO, but must be associated partially with the CuO$_{1-x}$ chains. Moreover, it is probably associated with the chain structural rearrangement that takes place at the ortho I-ortho II transition.

One possible resolution to these problems is to ascribe the neutron scattering in the doped samples near 50 meV (400 cm$^{-1}$) to modes associated with LO CuO(1) vibrations at chain ends. At a chain end because of the O(1) vacancy, the filamentary polaron must make a sharp turn onto an apical O(4) to continue its path. The infrared-active intact chain LO frequency[14] is 600 cm$^{-1}$, and at a softened chain end near 400 cm$^{-1}$ one would expect to find a state with broken LO symmetry (relative to the b-axis) due to back bonding of the chain-end Cu to the apical O(4). Thus the excess scattering seen in both the infrared and in the neutron spectra near 50 meV (400 cm$^{-1}$) would have a common origin, a II-III TO-LO hybrid with scattering strength enhanced by coupling to the filamentary current fluctuations that screen fluctuating internal electric fields.

It is important to realize that the filamentary structure shown in Fig. 7 in general is only a part (but the most important part) of the entire glassy structure. For example, in YBCO in addition to the semiconductive BaO layers between the cuprate planes and chains, there are also the Y planes between the chains, that contain no O in the ideal crystal structure. Even when Y is replaced by a strongly magnetic element, such as Eu or Gd, this has little effect on HTSC. From this one can infer that the filamentary paths generally avoid these non-oxide planes. However, for c-axis currents to flow, there must be some paths across the Y planes between the sandwiches shown in Fig. 7. It is known that when the samples are overdoped and contain regions with x = 7.0, the *c*-axis resistivity can be nearly linear in T, much like the *ab* planar resisitivity. This could be the result of some small O "interstitial" doping of the Y plane. In any case, while strong correlations are observed between the structure associated with chain ordering in the CuO$_{1-x}$ plane, there is no evidence that HTSC in YBCO is associated with the rare paths through the Y planes, and thus these planes are omitted from Fig. 7.

Another point one must keep in mind when examining data is that for any x one must always expect glassy compositional fluctuations from one region of the sample to another, even for "nearly perfect" single crystals as characterized by diffraction. In order to simplify the discussion, compositional trends are discussed as if only the dominant phase (metallic, insulating, or filamentary) is present at any given composition or dopant level x.

Because neither the nanodomain walls nor the dopants have long-range order, neither can be observed directly by diffraction, so that both of these key structural elements seemed to have the character of "hidden constructs", especially when they were first introduced[6]. By now, however, there is abundant evidence for the existence of such structural units in perovskites, with a domain diameter in the cuprates of 3-4nm. As this evidence is crucial to the model and is one of the origins of broken symmetry, we now discuss it in detail.

## V.     PARTITIONING OF $CUO_2$ PLANES INTO NANODOMAINS

Perovskite materials belong to a class of ferroelastic crystals[32]. A crystal is ferroelastic if it has two or more stable orientational states in the absence of a mechanical stress (and an electric field), and if it can be reproducibly transformed from one state to another of these states by the application of mechanical stress[33,34]. Reported maximum values of an atomic displacement in a variety of ferroelastic crystals range from 0.04 to 0.24 nm. A transformation from one orientational state to the other in ferroelastic crystals is in general accompanied by the formation of ferroelastic domains which reflect concentration of strain due to static correlated displacements of atoms from their periodic lattice sites. While it is easier to observe larger domain sizes, with modern experimental methods there has been a trend towards observation of smaller and smaller dimensions. For example, transmission electron microscopy (TEM) studies of a well know ferroelastic crystal of barium titanate ($BaTiO_3$) revealed the presence of domain structures of the size down to 4-6 nanometers[35]. The onset of ferroelasticity, as a function of temperature or pressure is often accompanied by additional cooperative phenomena such as ferroelectric or magnetic ordering. Correlation between the ferroelastic and superconducting transition has been also observed in conventional superconductors such as $V_3Si$ and $Nb_3Sn$[34].

One of the first experiments to have indicated a ferroelastic nature of HTSC cuprates was based on stress-strain measurements of $Bi_2Sr_2CaCu_2O_X$ whiskers of $T_c=75K$ by Tritt et al.[36]. <2212> whiskers are flexible ribbon-shaped single crystals with a c-axis (c=30.89 Å) perpendicular to the plane of the ribbon. The a-axis (a=5.414 Å) is the growth direction of the whiskers, perpendicular to the b axis (b=5.418 Å). Both a and b axes lie in the plane of the ribbon. The force (stress) was applied to the sample along the a-axis, and the displacement (strain) was measured capacitively. The results show hysteresis in the stress-strain curves above 270K, in addition to a maximum of Young's modulus at 270K. This suggests the existence of a structural phase transition. The data is characteristic of a displacive phase transformation and in particular of a ferroelastic transformation with the hysteresis which resembles that for a ferroelastic transition. The hysteresis and a temperature dependence of Young's modulus indicate a stress-related formation and dynamics (the hysteresis relaxation time is about 20 seconds) of ferroelastic domain walls.

MeV helium ion channeling has been used[37,38] to probe lattice distortions (static or dynamic) in $YBa_2Cu_3O_{7-\delta}$ as a function of temperature and oxygen doping. Ion channeling provides a direct real space probe of extremely small (sub-picometer) displacement of atoms in single crystalline materials. Sharma et al measured the excess lattice distortion above the thermal background $u_{ex}$ as a function of temperature between 30 and 300K. The magnitude of $u_{ex}$ was extracted from the measured FWHM (full-width-at-half-maximum) of the channeling angular scan. $u_{ex}$ exhibits cusps at several higher temperature ferroelastic transitions that are sample-dependent. Near $T_c$, $u_{ex}$ drops further with decreasing temperature.

Values of u can also be obtained from diffraction measurements (Debye-Waller factors), and these are compared to the ion-channeling results in Fig. 8. The diffraction values of u are based on the EMA, and correspond to an assumed Gaussian distribution of u values. The ion-channeling values of $u_{ex}$ measure non-planarity and weight large values of u much more heavily than the diffraction values, so that a few large values can dominate the measurements. Loosely speaking, one can say that diffraction measures $X = \langle u^2 \rangle$, while ion-channeling depends on averages of higher powers of u, such as $Y = \langle u^4 \rangle / X$. With a Gaussian distribution, $Y = 3X$, but with a Lorenzian distribution, Y diverges. Dechanneling events thus increase Y compared to X, and make dY/dT much more sensitive to reconstructive ferroelastic phase transitions than dX/dT.

The cusp that is observed at $T_c = 90K$ in Fig. 8 is, if anything, sharper than that observed for the ferroelastic transition at 180K, but the functional forms are similar. Both indicate the development of a nanodomain structure that can be described as a disorder-order transition with decreasing T that "locks in" below the phase transition. This behavior, with increasing constraints, can be compared to the stiffness transitions that occur in network glasses[8,10]. From the similarity of the ferroelastic and superconductive cusps, both as to functional form and magnitude, one must infer that ferroelastic effects involving nanodomain ordering are characteristic of HTSC.

Kaldis et al.[39,40] observed a displacive structural transformation in the copper-oxygen planes of $YBa_2Cu_3O_x$ at the optimally doped – overdoped phase separation line at x = 6.95. They measured, as a function of x, the dimpling in the $CuO_2$ planes using x-ray absorption fine-structure spectroscopy (EXAFS) at 25K and the oxygen O(2,3) in-phase mode Raman shifts. The data (Fig. 9) show for $x \geq 6.95$ anomalously large static displacements of the Cu(2) atoms off the O(2,3) layer (buckling), and a gap in the distribution of the O(2,3) in-phase Raman shifts. On doping from the underdoped side at x = 6.805 up to x = 6.885, the Cu(2) position was found to move smoothly along the c-axis by about 0.025 Å off the O(2,3) layer. From x = 6.895 up to x = 6.945 the dimpling of Cu(2) increases more rapidly by another 0.04 Å to its maximum value of 0.30 Å, almost entirely due to displacements of the Cu(2) atoms off the Y (yttrium) layer. At the onset of overdoping between x = 6.970 and x = 6.984 both the Cu(2) and O(2,3) layers shift off the Y-layer reducing the dimpling to 0.28 Å. Earlier work[40], based on standard refinements of neutron diffraction patterns (measured at 5K), has also shown a reduction in the dimpling of the CuO$_2$ planes by about 0.012 Å at x = 6.974.

The negative direction of this discontinuity has been attributed[40] to the structural transformation which develops first in small domains of the crystal. Anomalous softening of the O(2,3) Raman shifts, which starts at x larger than 6.90, has been found to correlate with the anomalously large displacements of the Cu(2) atoms off the O(2,3) layer observed by EXAFS around x = 6.95 (Fig. 10). The authors stated that the increase of the dimpling in the CuO$_2$ planes softens the Cu(2) – O(2,3) bonds and thus may decrease the wave number of the O(2,3) in-phase vibrations. The drop of the Raman shift by –5 cm$^{-1}$ within a narrow concentration range of $\Delta x \simeq 0.025$ gave evidence that the deformation of the CuO$_2$ planes is of the displacive type. On the other hand, an increase of the lattice constant in the c-axis direction is associated with a decrease of the a-axis lattice constant. The opposite behavior of these deformations suggests a martensitic (ferroelastic) nature of this phase transformation. Considering that the diffraction data average over domains and domain walls, and that most of the displacements are large only near the domain walls, the parallel between these phase transformations and those of network glasses[8,10] is very good. This reflects the fact that in both cases the driving force is stress accumulation between misfitting networks; the same mechanism is responsible for the formation of misfit dislocations at semiconductor heterointerfaces.

Pulsed neutron diffraction experiments[41] on YBa$_2$Cu$_3$O$_{6+x}$ (x =0.25, 0.45, 0.65, and 0.94) at 15K have revealed that the average copper Cu(2) - apical oxygen O(4) bond length changes from 2.4421 Å down to 2.2708 Å on going from z=0.1 up to z=0.94. The difference in these two bond lengths is 0.1713 Å. Displacements of Cu(2) in directions out of the a-b plane are also consistent with the presence of diffuse scattering in the electron diffraction which originates from the ferroelastic distortions[41].

One of the first studies that have addressed the presence of nanoscale structural perturbations (domains) in the copper-oxygen planes of cuprates, were the high resolution electron microscopy studies of YBCO by Etheridge[42]. They revealed a weak diffuse scattering in electron diffraction pattern of YBa$_2$Cu$_3$O$_{6+x}$ (x > 0.9), which arise from static displacements of atoms from their periodic lattice sites. The local atomic displacements partition the copper-oxygen planes into cells with dimensions comparable to a coherence length in the a-b planes (about 2 nm). These features were reported to be intrinsic to the orthorhombic form of $YBa_2Cu_3O_{6+x}$ (x > 0.9), and independent of the fabrication route. The intensity distribution of the diffuse streaks in the electron diffraction pattern is characteristic of scattering from atomic displacements and not of scattering from vacant chain-oxygen sites. It has been determined that there must be a static component to the atomic displacements, however the presence of a dynamic component to these displacements has not been found. In order to give the observed dark line contrast along axes perpendicular to c, the two-dimensional grids (cells) must be correlated along the c-axis (Fig. 10).

From selected-area electron diffraction patterns, static atomic displacements have been identified with the geometry consistent with displacements of the planar copper and apical oxygen atoms that are nearest neighbors along the displacement directions. High resolution images have revealed that there are connected networks of oxygen-pyramidal planes (which link the a-b planar oxygen atoms with the apical oxygen atom in the $CuO_5$ pyramid) at which the charge density distribution is locally perturbed. These effectively partition each of the copper-oxygen planes into "cells" with dimensions of about 2 nm in the a-b plane. These studies have suggested that the a-b planes of $YBa_2Cu_3O_{7-\delta}$ ($\delta < 0.1$) buckle into the network of slightly misaligned cells in a struggle to relieve internal stresses. One source of internal stress is the mismatch between "natural" (sometimes called prototypical) lattice constants of the copper-oxygen planes and the chain-oxygen and barium-oxygen planes ("natural" lattice constants are the ones that would minimize the energy of the plane if it were in isolation). The planar copper atom in the $CuO_5$ pyramid is considered to be tightly bonded in the center of the oxygen pyramid so that the pyramid might be treated as a rigid unit (Fig. 11). As the structure distorts in a response to internal strain it might be expected to yield most easily at the "soft" apical oxygen site, possibly causing the $CuO_5$ pyramid to tilt. These features suggest the effects of ferroelastic interactions.

Electrical transport and magnetic measurements of optimally doped YBCO have revealed[43] that the copper-oxygen planes behave like nanogranular superconducting systems (e.g. a conventional nanogranular niobium nitride (NbN) thin film). In a nanogranular superconductor with a coherence length comparable to the grain size, the temperature dependence of the critical current density $J_c$ is governed by the Josephson tunnel junctions at low temperatures where the Ginzburg-Landau (GL) coherence length is smaller that the grain size. At higher temperatures (close to $T_c$) where the GL-coherence length is larger than the grain size, the temperature dependence of $J_c$ is determined by the suppression of the order parameter in the grain. At the crossover temperature $T^*$, the GL coherence length equals[43] the grain size (for NbN nanogranular film $T^* \approx 0.85 T_c$). Consequently, at temperatures $T \leq T^*$, $J_c(T)$ is described by the Ambegaokar-Baratoff theory and at $T > T^*$ by the Ginzburg-Landau one with $J_c(T) \propto (T_c - T)^{3/2}$.

It has been found that $J_c(T)$ in YBCO exhibits this type of behavior[43] (Fig. 12). The measurements of $J_c(T)$ in the a-b planes of YBCO have been done on c-axis oriented thin film ring-shaped samples. This geometry allows one to determine $J_c(T)$ simply from the radial profile of the axial component of the self-field $B_z(r)$ of a maximum (critical) persistent current $I_c$. The relationship between $I_c$ and $B_z(r)$ is provided by the Biot-Savart law. The magnitude of the Ginzburg-Landau coherence length at $T^*$ has been used to estimate the size of the grain (cell) in the a-b planes to be approximately 3-4 nm. The cell size of 3-4 nm includes the width of the cell walls. This is in rough agreement with the TEM result[42].

The existence of large number of identical regions with diameters of about 3 nm (which have a relatively high density of low energy quasi-particle states) has been revealed[45] in $Bi_2Sr_2CaCu_2O_{8+\delta}$ ($T_c$=87K) using low-temperature (4.2K) scanning tunneling spectroscopy (STS). The studies have been carried out with a high resolution scanning tunneling microscope (STM) which simultaneously could measure, with atomic resolution, both the surface topography and the local density of states (LDOS) of a material. Following atomic resolution imaging of 130x130 nm² area, the mapping of the differential conductance at zero-bias was performed on the same area. This kind of map is a measure of the LDOS of low energy quasi-particles, and in a superconductor well below $T_c$, is expected in the EMA to show a very low differential conductance. In contrast to this expectation, a typical zero-bias conductance map of BSCCO revealed a large number of localized features (domains), with a diameter about 3 nm, which have a relatively high zero-bias conductance (high LDOS near the Fermi energy) (Fig.13) . These features appear to be randomly distributed. The authors suggested that these LDOS features are caused by quasi-particle scattering from atomic-scale defects or impurities, with oxygen inhomogeneities being most likely, due to the absence of the magnetic field effects. In view of the previously described evidence for ferroelastic nanodomains, the local high conductance features also could originate at randomly distributed highly conductive (intermediate phase) nanodomains, which are observed to have similar dimensions in other experiments.  The nanodomain diameters are observed to be constant, independent of doping, which is consistent with their origin as the result of ferroelastic misfit between the $CuO_2$ planes and the semiconductive planes (here SrO).

Scattering from nanodomain walls may also be the origin[46] of gigantic magnetoresistance (GR) effects which have been observed in manganite and nickelate perovskites.  Nanodomain structure has recently been observed[47] by electron microdiffraction from single-crystal regions of $La_{2/3}Ca_{1/3}MnO_3$. The diameter of the nanodomains in the orthorhombic plane is 3.6 nm, and their thickness normal to the planes is 1.5 nm. These nanodomain dimensions are very similar to those observed in the cuprates.

The "brick wall" nanodomain structure proposed in 1990 and shown in Fig. 7 has recently been realized in a context that is favorable for study by TEM. Alternating layers of a-Si and a-$SiO_2$ are annealed to produce nanocrystals (nc) of c-Si that form a nearly regular, self-organized, oriented array[48]. From the trends in nc size with film thicknesses, it is clear that the nc dimensions are determined by a combination of nucleation kinetics and self-limiting strain.  Although the mechanical aspects of nano-domain (-crystal) formation are similar in the two cases, the special feature that distinguishes the cuprates is the existence of an intermediate phase of a three-dimensional filamentary metallic nature over a narrow doping range.

## VI.     ASPECTS OF ZIGZAG FILAMENTARY MODEL

Because the structure and properties of HTSC are so complex, it seems almost inevitable that microscopic theories should develop as collections of metaphors, with different qualitative pictures (or, as some authorities would have it, dogmas) to describe each group of observables.  This is the safe and sure way to proceed, with independent (and not necessarily consistent) sets of adjustable *concepts* and adjustable *parameters* for each observable, and in the EMA it is unavoidable.  Here we have adopted a different approach, based on a unified topological picture of connectivity phase transitions.  We believe that the unified topological picture applies not only to HTSC, but also to the metal-insulator transition of uncompensated semiconductor impurity bands in the limit T $\rightarrow$ 0, to the temperature scaling of the resistivity of ultrathin metallic films, and to the stiffness transitions of network glasses: in other words, to connectivity transitions in any self-organized, topologically disordered network[7,8,49].

There have been remarkable strides in sample preparation recently, and it has become possible to prepare overdoped samples of $YBa_2Cu_3O_{6+x}$ , that we label as x = 7.00 samples, in contrast to optimally doped x = 6.93 samples[16,25,50].  The temperature dependence of the *c*-axis resistivity above $T_c$ in the x = 7.00 samples is a mixture of the linear dependence seen in optimally doped samples, and a small positive quadratic term that indicates an admixture of some Fermi liquid regions.  Similarly, the frequency dependence of the *c*-axis conductivity for x = 6.93 is almost constant, with a Drude upturn only at very low frequencies.  For x = 7.00 a marked Drude upturn (denoted by $\sigma_1$) is observed[16] for T = 100K below 300 cm$^{-1}$ in addition to the nearly constant non-Drude term $\sigma_2$.

In EMA language the nearly constant non-Drude term $\sigma_2$ is often described as a "hopping" term, and the different functional dependencies of  $\rho(T)$ and $\sigma(\omega)$ are explained[25] as "local field corrections" [of some unspecified kind]. These are excellent metaphors, as such, but with the filamentary model we can go much further, without introducing any metaphors. First we note that the linear dependence of ρ(T) for T > T′ [the pseudogap temperature], as well as its (pseudogap) dip for T < T′, for underdoped YBCO, was explained long ago[51]. In this model there is a high-mobility band pinned at $E_F$, which in the filamentary model is obtained by hybridizing chain and nanodomain plane states with resonant tunneling states (Fig. 7(a)). The dc resistivity is measured between electrodes, and so it is indifferent to the curvilinear character of these filaments: all that matters is the scattering that occurs as carrier wave packets traverse the filaments.

The situation is quite different for the infrared conductivity: here the probe is plane-wave photons, and the conductivity is measured by *projecting* curvilinear carrier wave packets onto these plane waves. When the wave packets make a right-angle π/2 c-axis interlayer turn through a dopant to avoid a vacancy on a chain, or a nanodomain wall in the plane, this looks to the plane-wave photon basis states like an admixture of Bloch states from energies over a range $E_F \pm \delta E$. In fact, knowing that the nanodomain diameter d is 3-4 nm, the planar unit cell dimension a is 0.5 nm, and the unit-cell band width W ~ 1 eV, we can estimate

$$\delta E \sim Wa/d \sim W/[6 - 8] \sim [0.16 - 0.12] \text{ eV} \qquad (1)$$

as the energy range over which the non-Drude "hopping" term $\sigma_2(\omega)$ is nearly constant. This number is in good agreement with the peak in $\rho_c(\omega)$ labeled A in the two-component model shown in Fig. 1 of ref. 25.

There is another way of formulating the "π/2 interlayer turn" effect. One can also consider the energy levels in the EMA or rigid band model, and ask how these are modified by the explicit presence [no premature averaging] of chain O vacancies, plane nanodomain walls, and resonant tunneling centers. What happens is that new states are formed near $E_F$ over a range $E_F \pm \delta E$. In other words, the energy levels in this range are strongly reordered, *not just renormalized with the same ordering, as occurs when the* *spatial inhomogeneities are treated by perturbation theory in the EMA.* This is why all perturbative field-theoretic EMA models of HTSC based on Fermi liquid theory are doomed to fail.

At this point the reader may think it appropriate for the paper to contain a large number of equations justifying these assertions. In fact, the authors believe that these assertions are nearly self-evident on the basis of the literature. For example, it has been well known for nearly 40 years that deflection of plane-wave photons by non-normal incidence on thin films[52] leads to breakdown of the forbidden character of LO phonon absorption for p polarization, as shown in Fig. 15. The magnitude of LO phonon absorption for an ideal film depends only on the ratio of the film thickness to the wave length of light, but in practice it is always enhanced by surface disorder. The deflection is treated classically in terms of boundary conditions at interfaces. The quantum-mechanical analogue is hybridization of electronic wave packets from different layers.

It has been more than 20 years since it was shown[53] that absorption by a mid-infrared defect in the perovskite $SrTiO_3$ at 163 meV is accompanied primarily by three one-LO phonon replicas, with the oscillator strength increasing in the replica series at 22, 59, and 100 meV. Thus the e-p interaction is much the largest for the LO phonons, and the LO-like $M_1^c$ electronic peak in YBCO is broad and strong just because it is associated with an interlayer defect (such as a chain O vacancy) that is an essential part of the filamentary path.

## VII. PHASE DIAGRAM FOR VIBRONIC $M_n^c$ PEAKS

It is convenient, from the point of view of connecting the electronic peaks to the phase diagram, to label the recently discovered[25] broad "A" peak in the *c*-axis conductivity for x = 7.00 as the $M_2^c$ peak. This is because this peak is weaker and broader for x = 6.93 and is therefore associated with the transition from the intermediate phase to the Fermi liquid phase. By the same token, the $M_1^c$ peak, although observed[15-20] for x = 6.5, 6.6 and 6.75, is associated with the transition from the insulating phase to the intermediate phase at x = 6.4. Its oscillator strength diminishes by a factor of two from x = 6.5 to x = 6.75, and it is no longer visible for x near 6.9.

According to the filamentary global topological model, the transition from the insulating phase to the intermediate phase at x = 6.4 is associated with the formation of 3-d filaments, as shown in Fig. 7. Such formation is associated with strengthening weak links that were essentially broken on the insulating side, where there were only filamentary segments. These weak links are primarily interlayer in nature, and thus involve the resonant tunneling centers. The latter are essential to the $M_1^c$ peak, which is thus the signature of these "defect" centers. The latter were mentioned in a very early paper[54] which also discussed pseudogaps. This phase transition is continuous.

There are two apparently contradictory quantitative chemical trends associated with the $M_1^c$ peak. With increasing x its energy increases, but at the same time its oscillator strength decreases. In models based on point-centered interactions, either short-range or Coulombic, increasing binding energy is always associated with increasing oscillator strength. The filamentary model accommodates the observed reverse correlation without introducing new assumptions or metaphors, as we will show.

Another problem is that it would seem that the density of tunneling centers (two per filamentary nanodomain) is too low to account for the large observed oscillator strengths. However, as one of us has pointed out elsewhere[49,55], the poor [non-metallic] screening in the semiconductive layer of resonant dopant potentials, and resonant dopant e-p interactions, produces an enhancement of these interactions by a factor of order (the number of unit cells)/nanodomain. This not only restores the oscillator strength to the level needed, but it also produces the giant e-p interactions needed to account for high $T_c$'s. Poor screening in the intermediate phase appears to be generic[51,55,56].

This interaction enhancement mechanism also explains the apparently contradictory chemical trends in energy and oscillator strength of the $M_1^c$ peak. The nanodomain array is glassy, which means that there will be a distribution of nanodomain sizes, even though the average size is nearly constant as a function of composition, as it is determined by the ferroelastic misfit between the $CuO_2$ layer and the adjacent semiconductive layer (in YBCO, the BaO layer). For low dopant densities near the metal-insulator transition, filaments will pass primarily through nanodomains with exactly two dopants and larger diameters n. Thus these filaments will see a larger $n^2$ enhancement factor and have larger oscillator strengths. As the dopant density increases, the larger nanodomains will be overdoped with more than two dopants and hence will be locally Fermi liquids; nanodomains with exactly two dopants will have smaller n, and filaments will pass through these smaller nanodomains with smaller $n^2$ enhancement factors. This leads to decreasing oscillator strengths with increasing x in $YBCO_x$.

What about the energy trends? In the presence of largely unscreened phonon interactions, one would expect a larger binding energy for longer longitudinal coherence lengths. An exact solution for a realistically curved and three-dimensional filament is not available, but in a model calculation larger LO phonon polaron binding energies were indeed obtained[57] for longer chains. The number of nanodomains/chain segment should increase as the dopant density increases to the point that the average nanodomain has exactly two electrically active dopants.

How was it possible for the filamentary model to reconcile two trends that were contradictory in point-centered models? The answer is that the filamentary model is a tensor description that contains two variable lengths, longitudinal and transverse, while point-centered models are scalar and contain only a single radial length. The oscillator strength depends primarily on *transverse* intradomain dimensions, and the vibronic excitation energy depends on the *longitudinal* interdomain coherence.

The second phase transition from the filamentary phase to the Fermi liquid is, in general, first order. The layered cuprates that form HTSC are complex multinary compounds, and preparing samples that are microscopically homogeneous, a necessary condition for observing first-order discontinuities, is not easy. Of course, unless such homogeneity is achieved, the second transition will be greatly broadened, and it will be difficult to show that the second transition is first-order. So far, at least three successful studies have reported evidence for first-order phase transitions: (1) near x = 0.21 in $La_{2-x}Sr_xCuO_4$ (after annealing for several months[58] at high T and constant O partial pressure), (2) near $\delta = 0.19$ in $Ba_2HgCuO_{4+\delta}$ (also after annealing at constant composition[59]), and near x = 0.95 in $YBa_2Cu_3O_{6+x}$ (carefully designed chemical and thermal history, including slow cooling[39]). Normally one observes only broadened trapezoids that appear to be parabolic $T_c(x)$'s. Self-organization is not easily achieved, and that is why only in a few experiments are the two HTSC transitions well separated to give sharply trapezoidal $T_c(x)$'s.

Here in the c-axis data on YBCO, the fact that the $M_2^c$ peak appears so abruptly[25] between x = 6.93 and x = 7.00, together with the fact that it is so strong for x = 7.00, are supporting evidence for the first-order nature of the transition between the filamentary phase and the Fermi liquid. In the latter the filaments have merged to form a liquid. In the intermediate phase there is only one filament per nanodomain. One way for filaments from different nanodomains to merge would be for carriers to be excited over nanodomain walls. One may therefore suppose that the height of the nanodomain walls has decreased abruptly to about 800 cm$^{-1}$ in the x = 7.00 sample, leading to the appearance of the $M_2^c$ peak at lower energies.

Although most of the focus of this paper is on YBCO and its alloys, we conclude this section on phase diagrams with Fig. 16 for $Ba_2HgCuO_{4+\delta}$. Normally one takes the oxygen doping concentration $\delta$ as independent variable, but in the recent diffraction studies[60] it has been found that the z spacing between O(2) and Ba in the BaO semiconductive plane gives more consistent results with less scatter. The importance of aplanarity in the BaO z spacing was also recognized in high pressure experiments[61]. The samples shown in Fig. 16 were prepared with a 300K-400K range of annealing temperatures, as well as a range of oxygen partial pressures. If $\delta$ is plotted as a function of the Ba-O(2) z spacing, or the unit cell volume, for example, it is found that it is not a single-valued function, but is instead re-entrant[59]. Moreover, the five points clustered together near the maximum of $T_c$ are spread out over a range of $\delta$ from 0.06 to 0.16.

In order to explain these anomalies it was suggested that there might be two distinguishable sites for $O_\delta$, one in the Hg plane and an out-of-plane interstitial site[59]. Filamentary theory suggests a different explanation. From Fig. 16 we see that as oxygen is doped into the Hg plane and the number and average length of filaments (including segments in the plane) increases, the unit cell volume and the Ba-O(2) z spacing shrink: the filaments are knitting together the structure, pulling it together more tightly, and ironing out the BaO aplanarity. This is not just a charge-transfer effect, because if it were, then δ would be a single-valued function of the Ba-O(2) spacing. Also, if it were just a charge-transfer effect, the unit cell volume would not show a strong first-order jump at the phase transition from the filamentary to the Fermi-liquid phase. Presumably the filaments are more strongly relaxed due to stronger electron-phonon interactions when the BaO plane is more nearly flat.

## VIII. DOPING DEPENDENCE OF OXYGEN PHONON OSCILLATOR STRENGTHS

The coherence factors that affect the strength of the $M_1^c$ and $M_2^c$ peaks so strongly also have equally striking, but less readily explained, effects on the temperature shifts and oscillator strengths of the prominent phonon peaks. The three strongest TO peaks are the O(1) peak [CuO chain mode] near 280 cm$^{-1}$, the O(2,3) [CuO$_2$ plane modes] peak near 320 cm$^{-1}$, and the O(4) [interplanar or apical O mode] peak near 570 cm$^{-1}$. These have been reported and discussed in detail[15], and the present discussion is an extension, within the context of the filamentary model, of that earlier discussion. For the reader's convenience these oscillator strengths ($f_i \sim \omega_{pi}^2$) are reproduced in Fig. 17 for YBa$_2$Cu$_3$O$_{6+x}$ compositions between x = 0.95 (slightly overdoped, chains nearly complete), and x = 0.50, near the metal-insulator transition.

The overall shift of 20% in the composition dependence of the oscillator strength of the 280 cm$^{-1}$ O(1) chain mode between x = 0.5 and x = 0.95 is easily explained[15] simply by the proportionality of the number of chain oxygens to x. However, one also notices in Fig. 17 a kink near x = 0.7 associated with the change in screening when the chain O's reorder. The composition dependencies of the oscillator strengths of the O(2,3) [CuO$_2$ plane modes] and the O(4) [interplanar or apical O mode] peaks are very much larger, however (factors of 3-4 overall), and were not discussed, as they would seem to be only distantly related to the chain site occupation. The point is that in the superconductive state the screening of the filamentary paths weakens as the gap opens on more and more paths, and the paths also pass through both O(2,3) and O(4) sites. The fact that the largest changes are seen for the latter can be explained if one assumes that carriers spend relatively more time on the O(2,3) and O(4) sites than they do on the O(1) sites. So long as the local superconductive gap is ≥ phonon frequency, changes in screening have large effects on the oscillator strengths.

An important feature of the filamentary model is that the filaments are self-organized to maximize the filamentary conductivity. Because the filaments pass successively through chain O(1)'s, resonant tunneling centers in the Ba layers (which should involve apical O(4)'s), and weakly buckled plane O(2,3)'s, all three of these oscillator strengths can be expected to vary sharply as screening is reduced below $T_c$. The largest temperature shift of peak frequency is that of the O(2,3) [$CuO_2$ plane modes] peak near 320 cm$^{-1}$, which softens dramatically below $T_c$, but the O(4) sites have equally large shifts in oscillator strength, again a filamentary effect. The composition dependence of this softening contains two interesting features. As expected, it is largest at x = 0.50, and it decreases abruptly near x = 0.7, the *chain ordering composition.*

As noted previously[15], in a simplistic EMA spring constant model, where the spring constants depend mainly on the interlayer c axis spacing, such sharp variations are not predicted, as nothing special happens to the c axis spacing itself at $T_c$. What is actually happening is that at $T_c$ the onset of superconductive order gives rise to sharp reconstructive effects associated with filamentary weak links. As the density of the latter is small (a few %), such reconstructive effects are not observable by diffraction, which is dominated by scattering from the large majority of only slightly reconstructed unit cells. However, the reconstruction near the weak links, although it involves only a small fraction of atoms, does involve just those atoms with large displacements from ideal sites. Scattering from the latter is very strong in ion channeling, and experiments on the latter have shown (Fig. 8) sharp effects on effective Debye-Waller factors at $T_c$ on the same samples where powder diffraction measurements of the actual Debye-Waller factors are smooth through $T_c$.

There are several ways that weak link reconstruction can occur, involving small atomic displacements near the weak links, that improve the already high mobility of carriers in states pinned in the narrow filamentary band at $E_F$. All of these involve strengthening of the filamentary path that passes though both the planes and the chains. This explains why it is possible for the temperature dependence of the O(2,3) [$CuO_2$ plane modes] peak near 320 cm$^{-1}$ to vary abruptly when the chains reorder.

We return to the temperature dependence of the O(2,3) [$CuO_2$ plane modes] peak near 320 cm$^{-1}$ and the electronic $M_1^c$ peak shown in Fig. 6. Samples with naturally shiny, unpolished surfaces show a large shift of the center of the $M_1^c$ peak from 410 cm$^{-1}$ (x = 0.5) to 480 cm$^{-1}$ (x = 0.75), associated with the knee in $T_c(x)$. This shift suggests that more hybridizing weight in the $M_1^c$ peak is given to chain LO modes (higher frequencies than the plane TO modes) as chain occupation increases. Such large shifts are not seen in samples with mechanically polished surfaces, which suggests that surface damage reduces the chain contribution by breaking longer chain segments into shorter ones.

## IX. EFFECTS OF ZINC DOPING

The oscillator strength of the vibronic $M_1^c$ peak in Y(Cu$_{1-x}$Zn$_x$)$_3$O$_{6.6}$ alloys disappears very rapidly[17] with increasing x, and drops to 0 between x = 0.006 and x = 0.013. Similar behavior occurs as a function of temperature as T goes above $T_c$. However, the effect of Zn doping is not merely thermal. Raising the temperature has only a small effect on the phonon bands, but increasing x has an anomalously large effect on both the apical O(1) band and the $CuO_2$ plane modes. Some effect on the latter was to be expected, as for low x the Zn replaces Cu in the $CuO_2$ planes. However, the observed magnitude of the O(2,3) effect, a line broadening of ~ 40 cm$^{-1}$, is very close to the T shift ~ 35 cm$^{-1}$ observed between T = 0 and T = $T_c$. Considering that the mass of Zn is nearly the same as that of Cu, such large shifts for a Zn concentration of ~ 1% cannot be explained with a classical spring model in the absence of internal structure in the $CuO_2$ planes. (See also the discussion of the neutron data in Fig. 3(b)).

It has been suggested[17] that these anomalies can be explained within a continuum model as the result of a change in antiferromagnetic spin interactions. However, spin interactions, like bonding forces, are short range, and would scarcely be likely to have much effect on vibrational line widths at these very low Zn concentrations. Instead, to explain these anomalies all one needs is a simple extension of the same ferroelastic effects that give rise to the nanodomains discussed in Sec. III. The natural site for the Zn in the $CuO_2$ planes is at the corners of the nanodomains, as the nanodomain walls function as relievers of the interplanar $CuO_2$-BaO ferroelastic misfit. Generally the bonding distortions at the domain walls must already be large, and so the strain energy will be minimized if the Zn occupies a more distorted wall site rather than an undistorted domain interior site. Moreover, if the Zn does even better, and occupies a corner site, it is likely to quench the local pseudogap associated with the neighborhood of that site, converting the associated nanodomains from coherent and filamentary regions, to Fermi liquid (effectvely overdoped) non-superconductive regions. Note that there are $\sim n^2$ unit cells/nanodomain, and that a corner Zn will affect four nanodomains. With $n \sim 6$, this means that $\sim 1/(4n^2) \sim 1\%$ Zn can quench filamentary vibronic anomalies, in agreement with experiment.

One would like to observe the Zn occupation of the corner sites directly. This seems to be a general topological property of the model and so it might have been observed[45] in Zn-doped $Bi_2Sr_2CaCu_2O_{8+\delta}$ ($T_c$=87K) using low-temperature (4.2K) scanning tunneling spectroscopy (STS). In fact, the authors note that in addition to the zero-bias bright spot associated with each Zn atom, one can discern narrow tetragonal spokes emanating up to 3 nm from a spot. These spokes are not observed in connection with impurity or defect X states that produce zero-bias bright spots in undoped BSCCO. They interpret these spokes as evidence for d-wave superconductivity, which is supposed to be a characteristic feature of the host, and so should be observed in connection with either Zn states or X states.

We believe that other evidence for bulk host d-wave superconductivity is dubious[62,63] at best, and that such a modified Fermi-liquid d-wave model is incompatible with the existence and character of the intermediate non-Fermi liquid phase (Figs. 1, 2 and 16). Thus we prefer to interpret these spokes as representing both the anisotropy of the pseudogap which is localized in the nanodomain walls, as well as the anisotropy of the Zn resonant charge density that is centered at the intersections of these walls. Then the difference between the STS images for the X states and the Zn states is understood by the Zn site preference for nanodomain corners.

In passing, it is worth noting that in the published images the spokes are quite narrow, and one would assume that the original images were even narrower. Their angular variation (on a logarithmic scale!) appears to be much too rapid to be described by a merely quadratic function, such as $F(x,y) = (x^2 - y^2)$, and a function such as $F^{2n}$, with $n \gg 1$, or even an exponential, would appear to give a much better fit. Where such rapidly varying angular terms would originate in a continuum model is difficult to imagine. However, in the nanodomain model the ratio (a/d) of nanodomain wall thickness a to nanodomain diameter d is expected to be ~ 1/6. This, together with the expected exponential attenuation of the Zn dopant zero-bias charge density outside the domain walls, would readily explain the observed narrowness of the tetragonal spokes.

One can now return to Figs. 3 (a) and (b), which contain a hidden anomaly. The 70 meV phonon is assigned[15] to apical O(4) modes, with its weak companion at 80 meV, seen strongly for $YBa_2Cu_3O_{6+x}$, assigned to O(4)'s connected to chain Cu(1)'s which have lost an O(1), in other words, chain ends. However, the $YBa_2(Cu_{0.9}Zn_{0.1})_3O_7$ data shown in Fig. 3(b) contain an 85 meV peak which has an integrated strength 1/3 to 1/2 as large as that seen for $YBa_2Cu_3O_{6+x}$. Thus each Zn must remove not only two nearest neighbor oxygens from the chain, but several oxygens from nearby Cu atoms as well. This seems to be consistent with the Zn corner site and the spokes observed by STS described above.

## X. COMPARISON WITH OTHER MODELS

A question that is often asked, but seldom answered, is "How does this theoretical model compare with other theoretical models?" We have already indicated, early and often, that all EMA continuum models designed to explain the $M_1^c$ vibronic peak, especially those that contain the magic metaphor "Josephson", seem inadequate to us. However, there are many constructive similarities between our discussion and the discussions[9,19] given by Timusk et al., even though their discussion modestly concluded by saying that they had "only the vaguest understanding of the nature of the phonons giving rise to the [$M_1^c$] band at 400cm$^{-1}$". In fact, their model is not so far from ours, providing that one re-interprets the coupling of an infrared inactive [LO] mode to a charge-transfer band, which they mention, not as just a simple (incoherent) charge-transfer band. Rather it is interlayer vibronic coupling through a resonant (and hence *coherent)* tunneling center pinned to $E_F$. It is the coherence of filamentary states that pass through the semiconductive layer to form pseudo-one-dimensional states, embedded in the three-dimensional maze created by planar nanodomains, that enables the model to explain giant electron-phonon couplings and so many internal structural features of *both* the neutron and the infrared data that appear to be contradictory in the EMA.

One of us has previously discussed[28] filamentary phonon coherence as a possible explanation for LO phonon dispersion anomalies in $La_{2-x}Sr_xCuO_4$ (x = 0.15) and $YBa_2Cu_3O_{6+x}$ (x = 0.20, 0.35, 0.60, and 0.93). The same neutron scattering data on large untwinned YBCO single crystals contain intensity anomalies along **q** = (y00) that cannot be explained[64] within the framework of the EMA. Similar dispersion anomalies have been observed in both superconductive ($Ba_{0.6}K_{0.4}BiO_3$) and ferromagnetic ($La_{1-x}Sr_x$ MnO) doped perovskites. These generic anomalies indicate that ferroelastic nanodomains are probably common to all doped perovskites that show very unusual physical properties.

If coherence is so important, then why isn't the Josephson plasmon metaphor adequate? Apart from the rather basic fact that neutrons do not couple to plasmons, there is a quantitative problem in explaining the origin of the giant e-p interactions.. The reason is that coherence alone is only one aspect of the problem, and when only that aspect is included within the EMA, the quantitative consequences of ignoring the internal network structure are fatal. Only a small fraction of the carriers (of order $T_c/W \sim 10^{-2}$, where W is the carrier band width $\sim$ 1 eV) is supposed to tunnel coherently, so that in the absence of internal network structure, the short-range e-p interaction in one unit cell will be so weak that the oscillator strength of the M peaks will be too small by a factor of $10^2$. In the filamentary model, the tunneling carriers in a given nanodomain are *funneled* into the unit cells of the tunneling centers. Thus localized, their charge density is large enough to create unscreened e-p interactions of the strength required to produce infrared absorption with the strengths even larger than that of TO phonons. The vibronic bands are broadened inhomogeneously because of the glassily disordered distribution of the tunneling centers. Finally, because the carriers are funneled through tunneling centers, where the LO c-axis phonon interactions are so large, the resulting electron-phonon interactions are indeed strong enough[6,7,51] to explain HTSC.

The filamentary nanodomain model is not merely a theory that explains neutron and infrared vibronic spectral anomalies. It is something much more general, a genuine theoretical platform that explains a wide range of anomalies. For example, it is instructive to compare the present topological model with the much more fashionable model of d-wave pairing. Andreev reflection at grain boundary interfaces shows structure that has been taken as evidence for d-wave pairing. The structure is predicted to exhibit a strong temperature dependence between $T = T_c$ and $T = 0$, but the structure is actually found to be temperature-independent[62]. In the topological model the structure is assumed to be associated with dopant filaments at the interface whose position is fixed, and hence must be temperature independent.

Even more striking are the results for nanoscale zero-bias defect or impurity states observed by scanning tunneling spectroscopy (STS)[45]. One considers two cases: native X defects, probably associated with O vacancies, and extrinsic Zn impurities. According to the d-wave model, the radial extent of the zero-bias states should depend on the nature or strength of the impurity (X or Zn) potential. On the other hand, the angular behavior should be characteristic of the host d-wave pairing itself and should be independent of the nature of the impurity. The nanodomain model predicts just the opposite: the radial extent of the zero-bias states depends only on the ferroelastically fixed nanodomain size of 3 nm, and is nearly independent of the nature of the impurity. The angular behavior is determined by whether or not the impurity is located within the domain, or at a domain wall or wall intersection. The X impurities are native and are most likely to be found in the domains, while the extrinsic Zn impurities are probably at the corner sites, because this is consistent with the known[65] critical doping level for Zn (about 1% replacement of Cu) and nanodomain size. The STS experimental data contradict d-wave pairing and appear to be in excellent agreement with the topological model[45].

Another feature of our nanodomain model, the association of pseudogaps with nanodomain walls, has been confirmed in recent STM data[66], which are again regrettably discussed in terms of "d wave superconductivity".

## XI.  CONCLUSIONS

The complexity, richness and variety of the many electronic, magnetic, and mechanical phenomena that have been observed in the cuprates is so great that many theorists have despaired of ever attaining a satisfactory theoretical model. Here we have argued that a satisfactory theoretical platform should begin with the strongest interactions, the ferroelastic ones. It is these interactions that generate the nanodomain pattern that, together with the layered structure, leads to the discrete, self-organized filamentary model illustrated in Fig. 7. This model is topologically unique. It explains chemical trends in observed properties that have the *wrong sign* in continuum models. It explains the drastic structural reorganization effects of a vibronic nature that have been observed in the density of states measured by neutron scattering, and quite similar anomalies in the infrared spectra (especially c-axis polarized), none of which are explicable in the EMA. It explains the large magnitude of doping effects (Cu(2) replaced by Zn) in YBCO, which are ~ 100x larger than can be explained with a continuum model. It shows that the numerous magnetic anomalies, although quite striking, are basically little more than an incidental by-product of the ferroelastic interactions. Magnetic interactions, although ubiquitous, cannot be the origin of HTSC, which is produced directly by Fermi energy electron-phonon local interactions at resonant tunneling centers, and indirectly by the ferroelastic geometry. This geometry itself is the result of non-local electron-phonon interactions, a different kind of interaction that involves all the electrons that contribute to interatomic misfit.

*Postscript*. After the completion of this manuscript, we learned of important work on very carefully prepared BSCCO junctions[67]. The tunnel junction characteristics dI/dV and $d^2I/dV^2$ are in excellent agreement with conventional Eliashberg theory, with the vibrational density of states measured by neutron scattering (analogous to Fig. 3 here), and with the vibrational density of states calculated by a lattice-dynamical model. Many peaks are observed, consistent with filamentary paths that closely resemble the ones discussed here (that is, involving $CuO_2$ and BaO planes, as well as apical oxygens and the very low frequency $Bi_2O_3$). The agreement between conventional s-wave theory (a very sharp 25 meV half gap isotropic to within 3 meV, and a predicted $T_c$ of 86(2) K) and experiment is probably better than has ever been attained before on any material. This truly remarkable work leaves little doubt that electron-phonon interactions cause HTSC. The contradictions between this work (which we believe to be correct in all details) and the many papers on "d wave pairing" are resolved in a detailed discussion that will be presented elsewhere.

# FIGURE CAPTIONS

Fig. 1. The connectivity or stiffness transition of network glasses is a function of the mean coordination number <r>. It also depends on whether the bonds are distributed randomly or are organized to minimize internal stress. The degree of internal connectivity is monitored by the fraction $f$ of floppy modes. As this figure shows, in a simple mean-field (Maxwell) model, $f$ is linear in <r> and vanishes at <r> = 2.40. In a numerical simulation with (nearly) random bonds, there is a single rigidity transition near <r> = 2.38. Finally, in a self-organized model, there are two transitions that define an *intermediate* phase. This *intermediate* phase is an excellent mechanical analogue of the "non-Fermi liquid" phase that is responsible for HTSC. So far it has proved impossible to obtain the intermediate phase in the absence of self-organization. For more information on these calculations, see ref. 10.

Fig. 2. The nature of the self-organization described in Fig. 1, and its effects on the second stress transition, are illustrated here, also from ref. 10. The second transition develops a pronounced first-order character when the stiff filaments are not allowed to close and form small rings; in this example, based on a diamond-lattice topology, the number of 6- and 8- membered rings, relative to the diamond lattice, is varied. These rings are analogous to small diamagnetic loops in the electromagnetic case. The exclusion of such small loops enhances the superconductive diamagnetic character of an electronic network, because the remaining loops are much larger, and more diamagnetic.

As the figure shows, the exclusion small loops makes the transition from optimally doped to overdoped more strongly first-order, as observed to be the case experimentally (see text and Fig. 16).

Fig. 3. Effective vibrational densities of states $G(\omega)$ measured by neutron scattering (ref. 11-13, 27). The decrease in intensity of the 40meV peak (TO O(2,3) plane) as O is added (a) to the O(4) chain site, together with the growth of a new peak at 50meV, not contained in vibrational models, is anomalous. Shell model calculations showed (ref. 11-13, 27) that these anomalies are inexplicable without the presence of a vibronic peak near 50 meV that borrows scattering intensity from the 40 meV peak, including a coherence factor. Similar coherence effects are needed to explain similar changes, also much too large, with addition of a few % Zn in (b).

Fig. 4. Growth of the vibronic peak labelled (?) near 400 cm$^{-1}$ = 50 meV in the c-axis infrared absorption (refs. 15,26), with oscillator strength borrowed from the 320 cm$^{-1}$ = 40 meV peak (TO O(2,3) plane), as a function of temperature. For this sample $T_c$ is near 58K. Approximately half of the growth takes place above $T_c$, which shows that the coherence effects are present in the normal state. According to the filamentary model, this reflects the coherence of filamentary quasiparticle states with energies below the nanodomain wall pseudogap. This coherence is a characteristic feature of the normal state, and it contributes to the dopant self-organization that is responsible for filamentary formation.

Fig. 5. The vibronic peak produces a correlation between the valley in the ab planar conductivity and the c-axis dielectric loss function, which is evidence for broken symmetry (ref. 26). In the filamentary theory this is explained by the sharp turn (see Fig. 7) in the filamentary path which occurs when it passes through a resonant tunneling center in a semiconductive layer, such as BaO in YBCO.

41Fig. 6. A more recent study (ref. 18) of the c-axis conductivity, showing a sharper rise [almost first-order] in the integrated strength of the 410 cm$^{-1}$ ~ 50 meV vibronic peak (filled squares) at $T_c$. The samples used in these experiments had as-grown, naturally shiny (unpolished) surfaces containing the c axis. Also shown is the *loss* of intensity of the 320 cm$^{-1}$ = 40 meV peak (TO O(2,3) plane) peak (open circles), which accounts for about half of the oscillator strength of the vibronic peak.

Fig. 7. The basic idea of the filamentary paths in the quantum percolative model (refs. 6, 7, and many others) for YBCO. The positions of the **I**nsulating **D**omain **W**alls in the $CuO_2$ layers are indicated, together with the **R**esonating **T**unneling **C**enters in the semiconductive layer, and oxygen vacancies in the $CuO_{1-x}$ chains. Giant e-p interactions are associated with the **R.T.C.**, where the interactions with LO c-axis phonons are especially large. The **I.D.W.** are perovskite-specific. The sharp bends in the filamentary paths are responsible for the broken symmetry that admixes ab planar background currents with c-axis LO phonons.

Fig. 8. The effective Debye-Waller factor displacement $U_1$ measured by ion channeling[37,38] is dominated by a few large displacements of atoms near oxygen vacancies[3] associated with resonant tunneling centers. These relax significantly near a ferroelastic domain transition, occurring in this YBCO sample near T = 180K, and also at $T_c$ = 90K. The effects of large displacements near $T_c$ are functionally similar to those near the ferroelastic transition, and this alone is evidence[3] for the importance of ferroelastic nanodomains for HTSC, as well as their strong interactions with superconductive filaments. The conventional Debye-Waller factor U, as measured by powder diffraction, weights all values of U equally (Gaussian weighting) and shows no effects at either transition, within the indicated error limits. Note that $U_1$ appears to be "frozen", in a glassy way, below $T_c$: this indicates that these large displacements contain

first-order components that are distributed over a range of temperatures because of fluctuations in local defect densities.

Fig. 9. Spacing between the Y-Cu(2) and Y-O(2,3) layers, as a function of oxygen concentration x in $YBa_2Cu_3O_x$ (see text). The spacing was determined from Y-EXAFS at 25K. Vertical arrows indicate the magnitude of dimpling in the $CuO_2$ planes. Here $x_{opt}$ marks the optimum oxygen concentration (From Kaldis et al Ref. 40).

Fig. 10. Raman shifts of the O(2,3) in-phase mode in $YBa_2Cu_3O_x$ for oxygen concentrations between x = 6.438 and 6.984. Dashed horizontal lines indicate the phase boundaries between coexisting phases, the drawn out horizontal lines (6.95) indicate the miscibility gap in the overdoped regime. The thick drawn boxes emphasize the sequence of phases occurring on doping (from Kaldis et al Ref. 40).

Fig. 11. A schematic illustrating the orientation of the irregular two-dimensional grid of walls of nanoscopic cells (approx. 2nm in size) in the $CuO_2$ planes of the $YBa_2Cu_3O_{6.95}$ (From Etheridge Ref. 42).

Fig. 12. The oxygen-pyramidal planes (delineated by the bold lines) in the average $YBa_2Cu_3O_{6.95}$ structure (From Etheridge Ref. 42).

Fig. 13. (a), (b), (c):Schematic representation of nanostructures in the a-b planes of $YBa_2Cu_3O_{7-\delta}$. The a- and b-axes are approx. 45 degrees relative to the domain walls, which are roughly along (110) directions (for details see Ref.43). An optimally doped superconductor could have nanostructures as in (a), where the critical current $I_c$ at low temperatures is governed by the interdomain Josephson junctions. The temperature dependence of $I_c$ is therefore that of Ambegaokar-Baratoff (AB) at low temperatures and that of Ginzburg-Landau (GL) above approx. $0.85T_c$ as described by the Clem's model (Ref.43) [(see solid triangles in (d)]. The dotted line in (d) shows $I_c(T)$ of a single Josephson junction (pure AB dependence). An underdoped superconductor is shown in

(b) where $I_c(T)$ is governed by the suppression of the order parameter in the nanodomain. $I_c(T)$ is governed by the GL dependence $[I_c(T) \propto (T_c-T)^{3/2}]$; see solid squares in (d). When a superconductor is a mixture of optimally doped and underdoped phases, its nanostructure is described schematically by "a disordered chessboard" shown in (c). In this case, $I_c(T)$ is expressed by open circles shown in (e). The solid line in (e) is a superposition of two components; an optimally doped one with an AB-like $I_c(T)$ (solid triangles) and an underdoped one with a GL-like $I_c(T)$ (dashed straight line). $\varepsilon_o$ in (d) and (e) is the Josephson coupling constant. See Reference 43 for more details.

Fig. 14. A 130nm square zero-bias conductance map (from Hudson et al Ref. 45). Quasi-particle scattering resonance (high LDOS) appear as bright regions approx. 3nm in diameter, owing to their higher zero-bias conductance. This size is independent of dopant concentration, and so is consistent with nanodomains whose size is determined by interlayer ferroelastic misfit.

Fig. 15. Transmitance of a thin (0.2 µ) film of LiF for non-normal incidence (about 30° from normal) in s and p polarization configurations (ref. 52). The deflection of the light induces absorption by LO phonons near 15µ, in addition to the symmetry-allowed TO absorption at 32.6µ. The strength of the LO absorption depends on the ratio of the (film thickness)/$2\pi$(light wavelength), here about $10^{-2}$. In the cuprates symmetry is broken much more strongly because the "film thickness" of a BaO semiconductive layer (for example) is less than $10^{-3}$ µ.

Fig. 16. In very carefully prepared samples of $Ba_2HgCuO_{4+\delta}$, the phase transition from the intermediate to the Fermi liquid phase is clearly first order (ref. 59-61). Overall there is much more scatter in $T_c(\delta)$ than in the indicated z BaO coordinate difference, which suggests the importance of Ba-O(2) planarity of the BaO plane, as opposed to the absolute doping level. This figure should be compared to Fig. 2, which is the only theoretical model so far to exhibit similar phase transitions.

Fig. 17. The composition dependencies of infrared oscillator strengths of the various TO phonon modes in YBCO$_{6+x}$, as a function of x, from ref. 15.

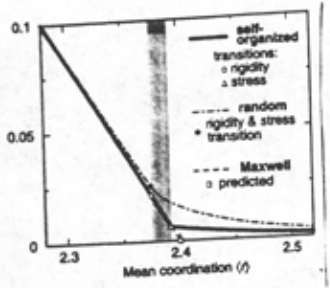

Fig. 1.

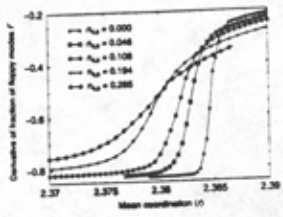

Fig. 2

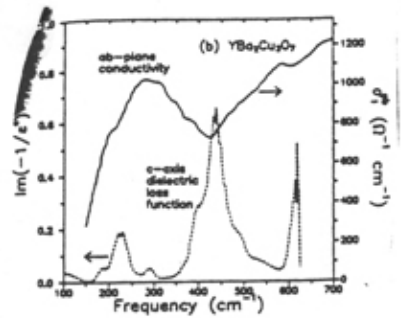

Fig. 5.

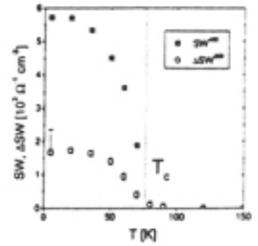

Fig. 6

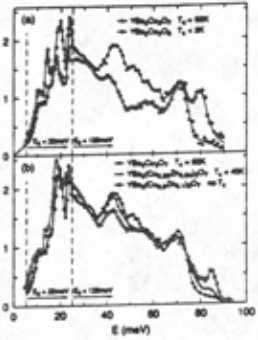

Fig. 3.

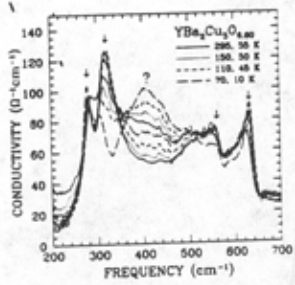

Fig. 4.

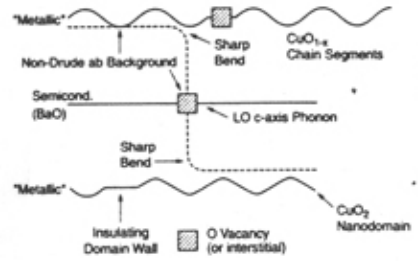

Fig. 7.

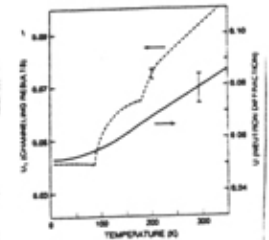

Fig. 8

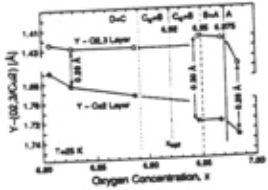

Fig. 9.

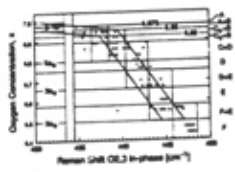

Fig. 10

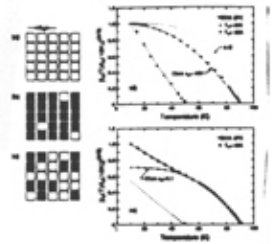

Fig. 13.

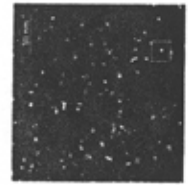

Fig. 14

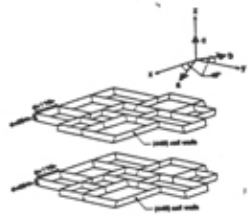

Fig. 11.

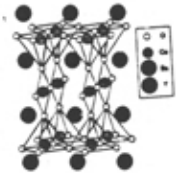

Fig. 12.

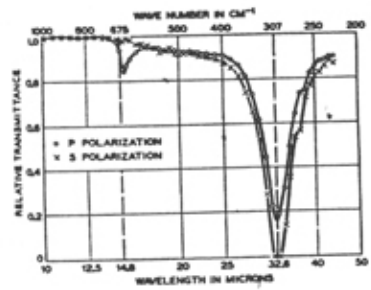

Fig. 15.

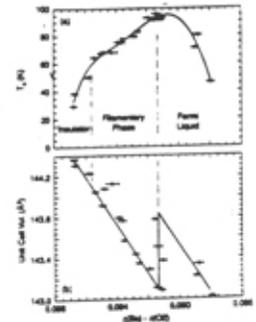

Fig. 16.

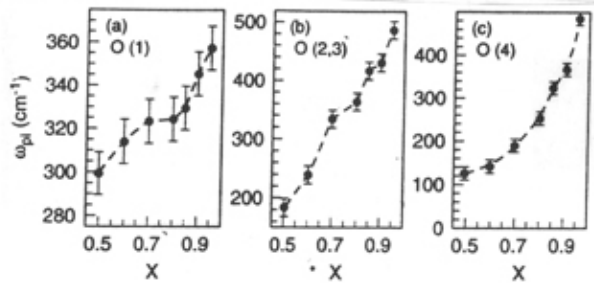

Fig. 17.